\documentclass{article}

\usepackage{arxiv}
\usepackage{tikz}
\usetikzlibrary{shapes.geometric, arrows}
\usetikzlibrary{positioning}
\usetikzlibrary{matrix,chains,decorations.pathreplacing,arrows}
\usepackage{float}
\usepackage{listings}
\definecolor{mygreen}{RGB}{28,172,0}
\definecolor{mylilas}{RGB}{170,55,241}
\usepackage{amssymb}
\usepackage{amsthm}
\usepackage[utf8]{inputenc}
\usepackage{tabularx}
\usepackage{tikz}
\usepackage{tikz-qtree}
\usetikzlibrary{mindmap,shadows}
\usepackage{url}
\usepackage{epstopdf}
\usepackage{etoolbox,ragged2e,siunitx}
\usepackage{array}
\usepackage{multirow}
\usepackage{media9}
\usepackage{amsmath}
\usepackage[title,titletoc]{appendix}
\usepackage{hyperref}

\title{Assessment of machine learning methods for state-to-state approaches}

\author{
  Lorenzo Campoli, Elena Kustova, Polina Maltseva \\
  Department of Fluid Mechanics, \\
  Saint Petersburg State University, \\
  7/9 Universitetskaya nab., \\
  St. Petersburg 199034, Russia \\
  \texttt{l.kampoli@spbu.ru}
}

\begin{document}
\maketitle

\begin{abstract}
It is well known that numerical simulations of high-speed reacting flows, in the framework of state-to-state formulations, are the most detailed but also often prohibitively computationally expensive. In this work, we start to investigate the possibilities offered by the use of machine learning methods for state-to-state approaches to alleviate such burden. 

In this regard, several tasks have been identified. Firstly, we assessed the potential of state-of-the-art data-driven regression models based on machine learning to predict the relaxation source terms which appear in the right-hand side of the state-to-state Euler system of equations for a one-dimensional reacting flow of a N$_2$/N binary mixture behind a plane shock wave. It is found that, by appropriately choosing the regressor and opportunely tuning its hyperparameters, it is possible to achieve accurate predictions compared to the full-scale state-to-state simulation in significantly shorter times. 

Secondly, we investigated different strategies to speed-up our in-house state-to-state solver by coupling it with the best-performing pre-trained machine learning algorithm. The embedding of machine learning methods into ordinary differential equations solvers may offer a speed-up of several orders of magnitude but some care should be paid for how and where such coupling is realized. Performances are found to be strongly dependent on the mutual nature of the interfaced codes.

Finally, we aimed at inferring the full solution of the state-to-state Euler system of equations by means of a deep neural network completely by-passing the use of the state-to-state solver while relying only on data. Promising results suggest that deep neural networks appear to be a viable technology also for these tasks. 
\end{abstract}



\section{Introduction} \label{sec:intro}
Various approaches to model strongly non-equilibrium
flows exist, such as the one-temperature (1T), multi-temperature (MT), and state-to-state (STS) approximations. The STS approximation, which assumes that characteristic times of vibrational energy transitions and chemical reactions are of the same order of magnitude as the gas-dynamic timescale, is the most detailed, since it can describe arbitrary vibrational energy distributions. For strongly non-equilibrium flows, the STS description, provides the most accurate results and the best agreement with experimental data~\cite{armenise2016advanced,kunova2015,kunova2016numerical,kunova2020}, nevertheless in many cases it is prohibitively computationally expensive. This led several researchers to investigate various energy binning approaches, both for CFD (Computational Fluid Dynamics) and DSMC (Direct Simulation Monte Carlo) simulation methods~\cite{magin2012coarse, munafo2014boltzmann, parsons2014modeling, torres2016uniform, berthelot2016modeling, sahai2016reduced, diomede2017insight}, as well as MPI-CUDA approaches \cite{bonelli2017mpi}.

In recent years, the increasing volumes of data, advances in computational hardware and reduced costs for computation, data storage and transfer, improvement of algorithms, an abundance of open source software and benchmark problems, and significant and ongoing investment by industry, led to an unparalleled surge of interest in the topic of machine learning (ML) (i.e., modern data-driven optimization and applied regression). Nowadays, ML algorithms are successfully employed for classification, regression,
clustering or dimensionality reduction tasks of large sets of
especially high-dimensional input data. In fact, ML has proved to have superhuman abilities in numerous fields and it is now rapidly making inroads also in fluid mechanics, providing a modular and agile modeling framework that can be tailored to address many challenges, such as experimental data processing, shape optimization, turbulence closure modeling, control and other traditionally intractable problems~\cite{schmidt2019recent,brunton2020machine,bruno2008transport,brunton2020special}

Given the growing interest in performing STS simulations and the availability of new investigation tools, in this paper, we start to explore the possibilities offered by the use of ML methods for STS approaches in order to reduce its computational cost.

\begin{figure}[H]
\centering
\scalebox{0.55}{
\begin{tikzpicture}[ every annotation/.style = {draw,
                     fill = white, font = \Large}]
  \path[mindmap,concept color=black!40,text=white,
    every node/.style={concept,circular drop shadow},
    root/.style = {concept color=black!40,
      font=\large\bfseries,text width=10em},
    level 1 concept/.append style={font=\Large\bfseries,
      sibling angle=50,text width=7.7em,
    level distance=15em,inner sep=0pt},
    level 2 concept/.append style={font=\bfseries,level distance=9em},
  ]
  node[root] {STS} [clockwise from=180]
    child[concept color=red] {
      node {\href{https://scikit-learn.org}{ML}}
    }
    child[concept color=blue, clockwise from=-5] {
      node[concept] {\href{http://texwelt.de}{Transport}} [clockwise from=65]
      child { node[concept] (brackets)
        {\href{http://texnique.fr}{Bracket integrals}} }
      child { node[concept] (coeffs)
        {\href{http://texwelt.de/wissen/}{Transport coeffs.}} }
      child { node[concept] (omega)
        {\href{http://texwelt.de/blog/}{Collision integrals} }}
      child { node[concept] (potentials)
        {\href{http://texwelt.de/blog/}{Intermolecular potentials} }}
      child { node[concept] (system)
        {\href{http://texwelt.de/blog/}{Transport linear systems} }}
    }
    child[concept color=green!40!black, clockwise from=145] {
      node[concept] {\href{http://texample.net/}{Kinetics}} [clockwise from=135]
      child { node[concept] (sections) {\href{http://texample.net/tikz/examples/}{Cross sections}} }
      child { node[concept] (rates) {\href{http://texample.net/tikz/examples/}{Rate coeffs.}} }
      child { node[concept] (relaxation)
        {\href{http://texample.net/weblog/}{Production terms}} }
      child { node[concept] (noneq)
        {\href{http://texample.net/community/}{Non-equilibrium factors}} }
    }
    ;
\end{tikzpicture}}
\caption{Machine learning for state-to-state: conceptual map.} 
\label{fig:concept}
\end{figure}
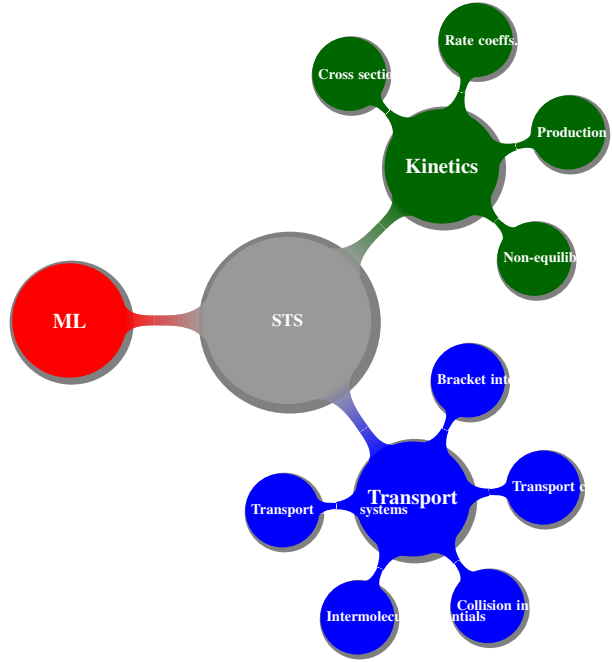

{In order to estimate the importance of different terms to the total computational cost, a simple overall profiling analysis was conducted for a typical two-dimensional simulation of an hypersonic non-equilibrium viscous reacting flow across a blunt body geometry. Details of input parameter settings and profile timings are given in Tab.~\ref{tab:example}. As expected the kinetic and transport modules are the most expensive. It is worth noting here that for the computation of the transport properties the Gupta~\cite{gupta1990review} model has been adopted and this is the reason why its cost is moderate. Nevertheless, as soon as state-to-state models are employed also for transport processes, the computational cost will be comparable or greater than the kinetic one.}
\begin{center}
\begin{table}[H]
\centering
\caption{\label{tab:example}Parameter settings and timings of one iteration (0.814s) for a typical two-dimensional simulation of an hypersonic non-equilibrium viscous reacting flow across a blunt body geometry.}
\begin{minipage}{2.55in}
\centering
\begin{tabular}{cc}
\hline 
Key & Value\tabularnewline
\hline 
\hline 
Flow Type & Navier\_Stokes\tabularnewline
Species & N2, O2, NO, N, O\tabularnewline
Kinetic & STS\tabularnewline
Transport & Gupta\tabularnewline
Gas Model & Nonequ\_Gas\tabularnewline
Simulation\_Type & 2D\_AXI\tabularnewline
Pressure & 2.0\tabularnewline
Temperature & 195.0\tabularnewline
Velocity & 11360.0\tabularnewline
Mass Fractions & N2:0.79, O2:0.21\tabularnewline
\hline 
\end{tabular}
\end{minipage}
\\
\vspace{.5cm}
\begin{minipage}{2.55in}
\centering
\begin{tabular}{cc}
\hline 
Process & Time\tabularnewline
\hline 
\hline 
Exchange & 0.00\%\tabularnewline
Update & 0.26\%\tabularnewline
ComputeConvectionFlux & 1.51\%\tabularnewline
ComputeDissipationFlux & 21.93\%\tabularnewline
ComputeSourceTerms & 76.27\%\tabularnewline
ComputeLocalResidual & 0.01\%\tabularnewline
\hline 
\end{tabular}   
\end{minipage}
\end{table}
\par\end{center}
{Consistently, it is possible to individuate several areas, showed in Fig.~\ref{fig:concept}, in which may be convenient to apply or at least to investigate the use of machine learning in the framework of state-to-state formulations. 
In the present paper, the Kinetic module will be considered while the Transport one will be addressed in a following publication.}
The paper is organised as follows: the mathematical formulation of the STS approach is first recalled in Section~\ref{sec:mathematics}. A detailed description can be found in~\cite{nk_book_e}. The regression of relaxation terms obtained by several ML algorithms is then presented in Section~\ref{sec:regression}. In Section~\ref{sec:coupling}, we couple a STS ordinary differential equations (ODE) solver and ML in order to speed-up the simulation by relieving the solver from the heavy computation of the kinetics source terms which are inferred by the ML. Section \ref{sec:EulerSTS} presents the inference of the full Euler system of equations for a one-dimensional reacting shock flow in the STS approach with a deep neural network (DNN). Finally, Section~\ref{sec:conclusion} summarises conclusions, open issues and future perspectives drawn from the present study.
\section{State-to-State mathematical formulation}
\label{sec:mathematics}
Experimental data relative to the relaxation times of different
processes in reacting mixtures~\cite{stupochenko1967relaxation} showed that in many cases of practical interest the following relation is valid:
\begin{equation}
\tau_{el}<\tau_{rot}\ll\tau_{vibr}<\tau_{react}\sim\theta\label{eq:-43}
\end{equation}
where $\tau_{el}$, $\tau_{rot}$, $\tau_{vibr}$, $\tau_{react}$ are respectively the mean times of translational, rotational, vibrational relaxation and chemical reactions and $\theta$ is the macroscopic gasdynamic time. Translational energy distribution is known to equilibrate fast and, for moderate temperatures, the rotational relaxation time is of the same order as the translational one and much smaller in comparison to the vibrational and chemical relaxation time. Therefore, processes of translational and rotational relaxation may be considered as rapid processes and on the contrary vibrational and chemical relaxation as the slow ones. The mean time of slow processes is comparable with the macroscopic time and these processes are strongly non-equilibrium. The condition given in Eq. (\ref{eq:-43}) provides the so-called state-to-state approach in non-equilibrium gas dynamics which describes the simultaneous processes of the vibrational and chemical relaxation. In this case, the macroscopic conservation equations for mass, momentum and total energy should be considered together with the equations for level populations of different chemical species since no quasi-stationary vibrational distributions exist.

A general state-to-state description of a non-equilibrium gas mixture flow can be found in~\cite{nk_book_e}. 
The master equations for the vibrational and chemical relaxation behind the shock wave in air~\cite{kn2014} include the 1D conservation equations of momentum and total energy
\begin{align}
\label{sys:momentum} & \rho v\displaystyle\frac{\partial v}{\partial x}+\frac{\partial p}{\partial x}=0,\\[1em]
\label{sys:energy} & v\displaystyle\frac{\partial E}{\partial x}+(E+p)\frac{\partial v}{\partial x}=0,
\end{align}
coupled in the frame of the state-to-state model to the equations for the vibrational state populations $n_{ci}$ of molecular species (N$_2$, O$_2$, NO)
\begin{equation}
\label{eq:nci} v\frac{\partial n_{ci}}{\partial x}+n_{ci}\frac{\partial v}{\partial x}=R_{ci}^{vibr}+R_{ci}^{react},\quad i=0,1,\ldots l_c,\quad c=1,2,\ldots l_m,
\end{equation}
and equations for the number densities $n_c$ of atomic species (N, O)
\begin{equation}
\label{eq:nc} v\frac{\partial n_{c}}{\partial x}+n_{c}\frac{\partial v}{\partial x}=R_{c}^{react},\quad c=1,2,\ldots l_a.
\end{equation}
In Eqs.~(\ref{sys:momentum})-(\ref{eq:nc}) $x$ is the distance from the shock front, $v$ is the gas velocity, $\rho$ is the mixture mass density, {$l_m$, $l_a$ are the numbers of molecular and atomic species, $l_c$ is the number of the upper vibrational state of molecule $c$ species,} $p$ is the pressure,  
$E$ is the total energy of the mixture per unit volume that can be presented as the sum of energies of translational, rotational, vibrational degrees of freedom and formation energy~\cite{nk_book_e}. {The translational and rotational energies are calculated on the basis of the local equilibrium Maxwell--Boltzmann distributions and are functions of temperature and chemical species number densities; the vibrational energy depends on the non-equilibrium populations of vibrational states. Thus, the calorically-perfect gas model is not applicable in the STS approach.}

The production terms $R_{ci}^{react}$, $R_{ci}^{vibr}$, $R_c^{react}$ describe the variation of vibrational level populations or mixture component number density due to the non-equilibrium kinetic processes such as chemical reactions and vibrational energy transitions: 
\begin{equation}
R_{ci}=R_{ci}^{vibr}+R_{ci}^{react}=R_{ci}^{VT}+R_{ci}^{VV}+R_{ci}^{2\rightleftharpoons2}+R_{ci}^{2\rightleftharpoons3}.\label{eq:-36}
\end{equation}
The expressions for the state-specific production terms are as follows:%
\begin{equation}
R_{ci}^{VT}=\sum_{M}n_{M}\sum_{i'\neq i}\left(n_{ci'}k_{c,i'i}^{M}-n_{ci}k_{c,ii'}^{M}\right),\label{eq:-35}
\end{equation}
\begin{equation}
R_{ci}^{VV}=\sum_{dki'k'}\left(n_{ci'}n_{dk'}k_{c,i'i}^{d,k'k}-n_{ci}n_{dk}k_{c,ii'}^{d,kk'}\right),\label{eq:-34}
\end{equation}
\begin{equation}
R_{ci}^{2\rightleftharpoons2}=\sum_{dc'd'}\sum_{ki'k'}\left(n_{c'i'}n_{d'k'}k_{c'i',ci}^{d'k',dk}-n_{ci}n_{dk}k_{ci,c'i'}^{dk,d'k'}\right),\label{eq:-28}
\end{equation}
\begin{equation}
R_{ci}^{2\rightleftharpoons3}=\sum_{M}n_{M}\left(n_{c'}n_{f'}k_{rec,ci}^{M}-n_{ci}k_{ci,diss}^{M}\right),\label{eq:-29}
\end{equation}
{$M$ stands for the collision partner which does not change its internal state during the collision; $k_{ii'}^{M}$, $k_{c,ii'}^{d,kk'}$, $k_{ci,c'i'}^{dk,d'k'}$, $k_{ci,diss}^{M}$, $k_{rec,ci}^{M}$ are the state-specific rate coefficients of vibrational energy transitions and chemical reactions. For the five-component air mixture,}
the following processes are included to the kinetic scheme: VV vibrational energy exchanges within the same chemical species; VV$'$ vibrational transitions between molecules of different species; single-quantum VT vibration-translation energy exchanges; all kinds of state-resolved dissociation reactions; Zeldovich exchange reactions taking into account vibrational excitation of both reagents and products. Thus, the list of reactions reads:
\begin{align}
\label{eq:vt}    {\rm VT}&: AB(i)+M \rightleftarrows AB(i\pm 1;i\pm 2)+M,\\[1em]
\label{eq:vv}    {\rm VV} &: AB(i)+AB(k) \rightleftarrows AB(i\pm 1)+AB(k\mp 1), \\[1em]
\label{eq:vvs}    {\rm VV'} &: AB(i)+CD(k) \rightleftarrows AB(i\pm 1)+CD(k\mp 1), \\[1em]
\label{eq:dr}    {\rm DR}&: AB(i)+M \rightleftarrows A+B+M,\\[1em]
\nonumber & AB, \, CD = {\rm N_2,\,O_2,\,NO},\quad M = {\rm N_2,\,O_2,\,NO,\,N,\,O},\\[1em]
\label{eq:er1}    {\rm ER}_1&: {\rm N_2}(i)+{\rm O} \rightleftarrows {\rm NO}(i')+{\rm N},\\[1em]
\label{eq:er2}    {\rm ER}_2&: {\rm O_2}(i)+{\rm N} \rightleftarrows {\rm NO}(i')+{\rm O}.
\end{align}
Molecular vibrational energy levels are calculated according to the anharmonic oscillator models. The total numbers of excited states are 122 and include 47 states of N$_2$, 36 of O$_2$, and 39 of NO. {For binary mixtures, we keep in the kinetic scheme only processes (\ref{eq:vt}), (\ref{eq:vv}), and (\ref{eq:dr}).} 
The relations connecting the rate coefficients of forward and backward collisional processes follow from the microscopic detailed balance after averaging with the Maxwell-Boltzmann distribution over the velocity and rotational energy~\cite{nk_book_e}. Accordingly, the rate coefficients of forward and backward vibrational energy transitions satisfy the relation: 
\begin{equation}
k_{c,i'i}^{d,k'k}=k_{c,ii'}^{d,kk'}\frac{s_{i}^{c}s_{k}^{d}}{s_{i'}^{c}s_{k'}^{d}}\frac{Z_{ci}^{rot}Z_{dk}^{rot}}{Z_{ci'}^{rot}Z_{dk'}^{rot}}\exp\left(\frac{\varepsilon_{i'}^{c}+\varepsilon_{k'}^{d}-\varepsilon_{i}^{c}-\varepsilon_{k}^{d}}{kT}\right)\label{eq:-30}
\end{equation}
where $s_{i}^{c}$ are the vibrational statistical weights, which for diatomic species are equal to 1. {$Z_{ci}^{rot}$ and $\varepsilon_{i}^{c}$ the rotational partition function and the vibrational energy of species $c$ and vibrational level $i$, respectively, $k$ is the Boltzmann constant, $T$ is the temperature,  and the prime denotes the energy levels of particles after a collision. Note that in the STS model, the rotational partition function depends on the vibrational state.} 

Similarly, for the chemically reactive collisions we can define the rate coefficients for exchange reactions $k_{c'i',ci}^{d'k',dk}$: 
\begin{multline}
k_{c'i',ci}^{d'k',dk}=k_{ci,c'i'}^{dk,d'k'}\left(\frac{m_{c}m_{d}}{m_{c'}m_{f'}}\right)^{\frac{3}{2}}\frac{Z_{ci}^{rot}Z_{dk}^{rot}}{Z_{ci'}^{rot}Z_{dk'}^{rot}} exp\left(\frac{\varepsilon_{i'}^{c'}+\varepsilon_{k'}^{d'}-\varepsilon_{i}^{c}-\varepsilon_{k}^{d}}{kT}\right)\label{eq:-31}
\exp\left(\frac{D_{c}+D_{d}-D_{c'}-D_{d'}}{kT}\right)
\end{multline}
and the rate coefficients for dissociation and recombination $k_{rec,ci}^{d}$:
\begin{equation}
k_{rec,ci}^{d}=k_{ci,diss}^{d}\left(\frac{m_{c'}+m_{f'}}{m_{c'}m_{f'}}\right)^{\frac{3}{2}}h^{3}\left(2\pi kT\right)^{-\frac{3}{2}}Z_{ci}^{rot}\exp\left(-\frac{\varepsilon_{i}^{c}-D_{c}}{kT}\right)\label{eq:-32}
\end{equation}
Here, $D_{c}$ and $D_{d}$ represent the dissociation energy of the molecule $c$ and $d$, $m$ the mass of colliding particles, $h$ the Planck's constant.

Adequate models for state-resolved rate coefficients of chemical reactions and energy transitions are crucial for accurate predictions of macroscopic flow variables. Detailed comparisons between various models can be found in~\cite{campoli2020models} and it is out of the scope of the present paper. 
In this study, we have used the following models:

\begin{itemize}
    \item \textbf{Vibrational energy exchanges}.
    Rate coefficients of vibrational energy transitions for processes (\ref{eq:vt})-(\ref{eq:vvs}) are calculated according to two models:
    \begin{enumerate}
        \item The Schwartz-Slawsky-Herzfeld (SSH) theory~\cite{ssh1952, hl2013} for description of VT and VV transitions of vibrational energy. This model was supplemented by the relation obtained on the basis of experimental data for VV$'$ exchanges of N$_2$-O$_2$ interaction~\cite{cfgo2013} (interaction with NO molecules was not considered in this case). This model was used for the binary mixture testcase.
        \item The Forced Harmonic Oscillator (FHO) model~\cite{amrt1998}. This model was used for the air mixture testcase.
    \end{enumerate}
    \item \textbf{Dissociation and recombination reactions.}
    State-specific dissociation and recombination reactions (\ref{eq:dr}) are described using the preferential Marrone--Treanor model~\cite{mt1963} which provides expressions for the state-dependent rate coefficients of dissociation in terms of the thermal equilibrium reaction rate coefficient and non-equilibrium factor depending on the parameter $U$ which characterises increasing the dissociation probability for the vibrationally excited states. Thermal equilibrium rate coefficients are commonly calculated using the Arrhenius law with the parameters extracted from the experimental data; the parameter $U$ can be found using the quasi-classical trajectory (QCT) simulations, see~\cite{KKS_cpl_2016}. Recombination rate coefficients are calculated using the detailed balance principle \cite{nk_book_e}. The following sets of parameters are used for the simulations:
    \begin{enumerate}
        \item The parameters in the Arrhenius law provided by Park~\cite{park1993}
        \item The parameter $U$ of preferential dissociation: $U = D/6k$ ($D$ is the species dissociation energy, $k$ is the Boltzmann constant)
    \end{enumerate}
    \item \textbf{Exchange chemical reactions.}
    State-resolved rate coefficients of exchange chemical reactions (\ref{eq:er1})-(\ref{eq:er2}) are calculated using the recently developed model~\cite{ksk2018} representing a modification of the Aliat model~\cite{aliat2008} improved by taking into account the vibrational states of NO and adjusted by comparison with the results of QCT calculations.
\end{itemize}
It is interesting to observe that each of the aforementioned model is characterized by its own computational efficiency which can significantly affect the overall time-to-simulation. Machine learning methods are agnostic respect to this aspect as they provide approximately the same efficiency independently of the model or the processes involved.
\section{Regression}\label{sec:regression}
Generally speaking, machine learning algorithms may be categorized into supervised, semi-supervised, unsupervised learning depending on the degree to which external supervisory information is available to the learning machine~\cite{brunton2020machine}. Supervised machine learning is based on the same principles as a standard fitting procedure: it tries to find the unknown function that connects known inputs to unknown outputs. The desired result for unknown domains is estimated based on the interpolation or extrapolation of patterns found in the labeled training data.

Classification and regression are the two most popular supervised problems. Whereas the outputs of classification are discrete class labels, regression is concerned with the prediction of continuous quantities~\cite{williams2006gaussian, stulp2015many, kostopoulos2018semi}.

The present section deals with the regression of relaxation terms, Eq.~\ref{eq:-36}, defined according to the STS formulation. In this regard, several state-of-the-art ML algorithms from the scikit-learn~\cite{pedregosa2011scikit} framework, reported in Tab.~\ref{tab:methods}, specifically, Kernel Ridge (KR), Support Vector Machines (SVM), k-Nearest Neighbor (kNN), Gaussian Processes (GP), several ensemble methods (Random Forest (RF), Extremely Randomized Trees (ET), Gradient Boosting (GB), Histogram-Based Gradient Boosting (HGB), Multi-layer Perceptron (MLP), were evaluated. Table~\ref{tab:methods} also reports in bold font the optimal parameters found by the grid-search analysis.

The dataset for this task was generated by an in-house Matlab code for the solution of STS one-dimensional flow relaxation behind a shock wave. Detailed description and results of the code can be found in~\cite{campoli2020models}. The dataset contains the molecular and atomic relaxation terms as functions of the distance from the shock, molecular and atomic number densities, velocity and temperature. 
The dataset was divided in the following way: 75\% of samples were used for training, whereas the remaining 25\% were used for testing, using the \texttt{train\_test\_split} built-in function. Hyperparameters are parameters that are not directly learned within estimators. They are passed as arguments to the constructor of the estimator classes. It is possible and recommended to search the hyperparameter space for the best cross-validation score. We used \texttt{GridSearchCV} to tune the hyperparameters. Each method was run with a preset grid of input parameters, detailed in Tab.~\ref{tab:methods}. The optimal setting of the parameters was determined based on 10-fold cross-validation performed on training data only. 

The algorithms were fed with scaled data. It is generally regarded as a good practice, to scale the input data to allow the model to more easily train and converge. Scaling the output targets also reduces the range of the output predictions, make easier and faster to train the network and enable the model to obtain better results as well. Moreover, data leakage is another issue for ML, that occurs when information from outside the training dataset is used to create the model. This additional information can allow the model to learn or know something that it otherwise would not know and in turn invalidate the estimated performance of the model being constructed. Such issue was avoided by using \texttt{fit\_transform} on the train data to learn and train the scaling parameters while only using \texttt{transform} on the test data with the scaling parameters learned on the train data. An assessment of the influence of model stability and performance on data scaling was conducted. Comparable results were found by using standardized (\texttt{StandardScaler}) and normalized (\texttt{MinMaxScaler}) approaches while unscaled data introduced severe issues in the algorithmic convergence. Nevertheless, the choice of the optimal preprocessing method for the particular regressor is out of scope of this paper.
\label{subsec:relax}

Table~\ref{tab:relaxationterm} compares the aforementioned ML algorithms for the regression of the relaxation terms. The mean absolute error (MAE), mean squared error (MSE) and root mean squared error (RMSE) have been used as metrics to evaluate the quality of a model's predictions. In addition, the coefficient of determination, $R^2$, regression score function is shown, as well as the training and prediction times. It is important to observe that these times as well as all simulations were run serially in order to have clean baseline estimates although it would be easy and advisable to take advantage of parallel processing (\texttt{n\_jobs=-1}).

The present results refer to a single-input, single-output regression. In other words, we estimate the ML algorithm on a single vibrational level for each relaxation term as a function of temperature. Once the "optimal" set of hyperparameters have been found through the aforementioned cross-validation procedure, the re-trained network is used to perform a \texttt{Multioutput} regression, (a simple strategy to extend regressors that do not natively support multi-target regression), to predict all the relaxation terms at once. 

\begin{table}[H]
\caption{Comparison of several MLAs for regression of relaxation terms.}\label{tab:relaxationterm}
\begin{centering}
\footnotesize{
\begin{tabular}{ccccccc}
\hline
\hline 
 Algorithm & MAE & MSE & RMSE & $R^2$ & $T_{train}\left[s\right]$ & $T_{predict}\left[s\right]$\tabularnewline
\hline 
\hline 
KR & 7.868505e-08 & 3.800217e-14 & 1.949414e-07 & 0.999999 & 7.612628 & 0.075077\tabularnewline
\hline 
SVM & 1.236652e-02 & 2.109761e-04 & 1.452501e-02 & 0.999786 & 5.317098 & 0.008577\tabularnewline
\hline
kNN & 8.655485e-04 & 2.659352e-06 & 1.630752e-03 & 0.999997 & 0.002296 & 0.004962\tabularnewline
\hline
GP & 7.235743e-07 & 2.436803e-12 & 1.561026e-06 & 0.999994 & 118.3911 & 0.098444 \tabularnewline
\hline 
DT & 2.417524e-03 & 1.623255e-05 & 4.028964e-03 & 0.999983 & 0.003520 & 0.000317 \tabularnewline
\hline 
RF & 1.140677e-03 & 5.016757e-06 & 2.239812e-03 & 0.999992 & 4.362630 & 0.038143\tabularnewline
\hline 
ET & 1.595557e-03 & 6.005923e-06 & 2.450698e-03 & 0.999993 & 2.279543 & 0.202767\tabularnewline
\hline 
GB & 2.300499e-03 & 1.478234e-05 & 3.844782e-03 & 0.999985 & 4.823793 & 0.006213 \tabularnewline
\hline 
HGB & 6.098571e-03 & 1.395461e-04 & 1.181296e-02 & 0.999859 & 14.385128 & 0.042188 \tabularnewline
\hline 
MLP & 6.023895e-03 & 7.539429e-05 & 8.682989e-03 & 0.999943 & 11.322764 & 0.009778 \tabularnewline
\hline 
\hline 
\end{tabular}
}
\par\end{centering}
\end{table}

By observing Table~\ref{tab:relaxationterm}, it is worth noting that the $R^2$ score parameter is not a good representative of the prediction quality since it reaches very high and similar values for all models. Conversely, the error metrics are more reliable discriminants. It can be observed that the Kernel Ridge algorithm reports the best results in terms of error levels while the Support Vector Machine, the worst. The remaining algorithms show comparable error levels, nevertheless, appreciable differences in the prediction time are noticeable. In this regard, the Decision Tree appears to be the fastest predictor. The k-Nearest Neighbour algorithm offers a relatively small prediction time, more than an order of magnitude faster than Kernel Ridge, with error levels second only to Kernel Ridge.

\begin{figure}[H]
\centering
\includegraphics[scale=0.5]{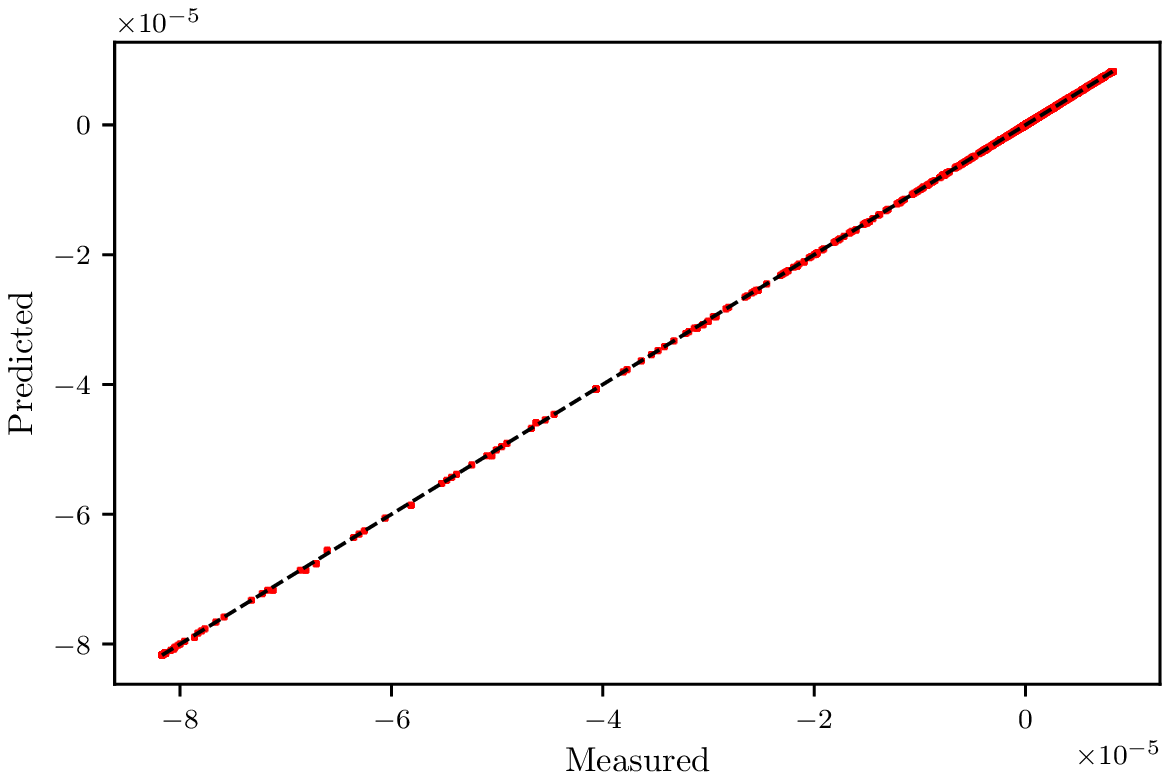}
\includegraphics[scale=0.475]{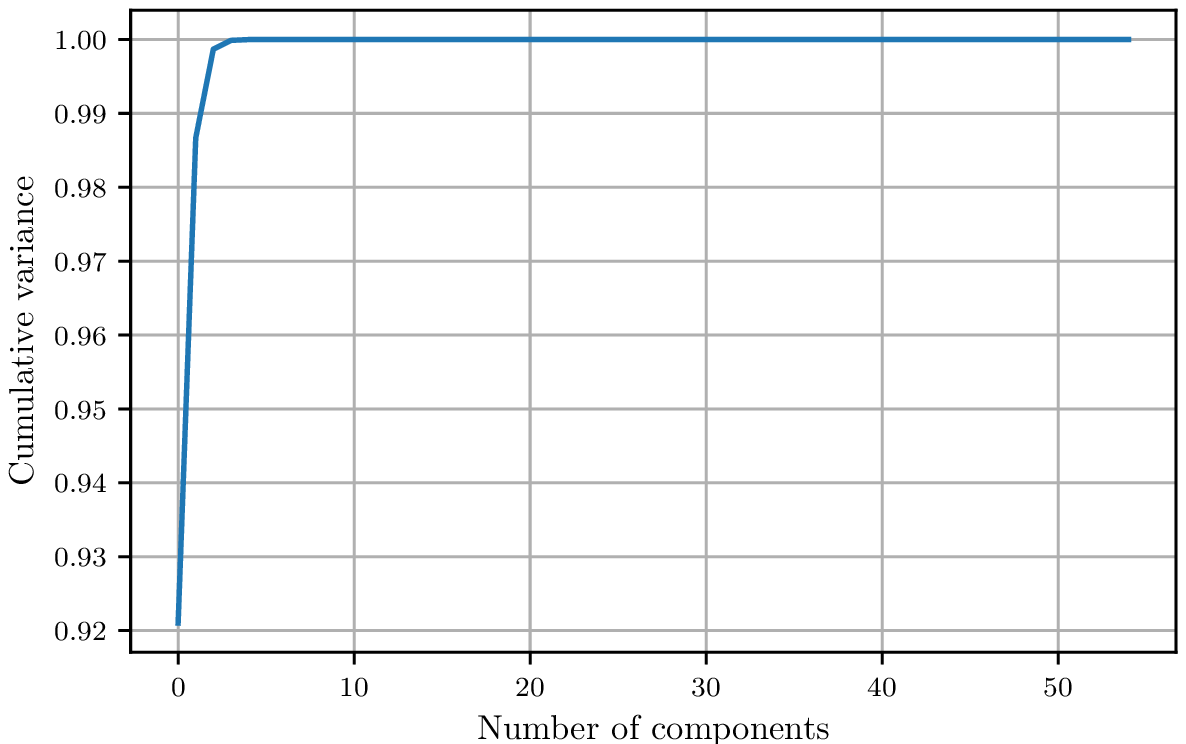}
\caption{Parity plot of cross-validated predictions against "ground truth" values for the Decision Tree algorithm (left) and plot of cumulative variance as a function of the PCA number of components (right).}
\label{fig:mparity}
\end{figure}

As a supplementary check of model's results, parity plots are presented in Fig.~\ref{fig:mparity} to show the closeness between predictions and actual ground truth values. Generally, a good spread is required as skewness of the points to a side of the diagonal would indicate a defective model. The only other parameter that can catch such a model flaw is the mean bias error (MBE) which is an aggregate value. In this study, parity plots were chosen as it is easier to appreciate and it gives a point by point information of which has been under-predicted or over-predicted. Even if, for the sake of clarity, only the Decision Tree results are shown here, all the presented algorithms perform similarly, with data points lying close to the diagonals and evenly distributed above and below the diagonal.
The extraction of flow features from experimental data and large scale simulations is a cornerstone for flow modeling. Moreover identifying lower dimensional representations for high-dimensional data can be used as pre-processing for all tasks in supervised learning algorithms. This task may be remarkably important in the framework of STS formulations which are currently heavily affected by the "curse of dimensionality". Consequently, in the present paper, we also investigated the possibility of dimensionality reduction by using the classical proper orthogonal decomposition (POD) or linear principal components analysis (PCA) algorithm which is a fast and flexible unsupervised method. Using PCA for dimensionality reduction involves zeroing out one or more of the smallest principal components, resulting in a lower-dimensional projection of the data that preserves the maximal data variance.
Specifically, in Fig.~\ref{fig:mparity} the cumulative variance as a function of the number of components clearly shows that by considering only 3 features, approximately the 98\% of the original data (made up of 100 input features) information is retained. 
\begin{table}[H]
\begin{centering}
\caption{Hyperparameters settings. The parameters in quotations refer to scikit-learn names. The parameters in bold font refer to the optimal values respect to the considered task.}
\label{tab:methods}
\footnotesize{
\begin{tabular}{ccc}
\hline 
Algorithm & Parameter & Values \tabularnewline
\hline 
\hline 
\multirow{3}{*}{KR} & kernel & \{poly, \textbf{rbf}\}\tabularnewline
\cline{2-3}
 & alpha & \{1e-3, 1e-2, 1e-1, \textbf{1e0}, 1e1, 1e2, 1e3\}\tabularnewline
\cline{2-3}  
 & gamma & \{1e-3, 1e-2, 1e-1, 1e0, \textbf{1e1}, 1e2, 1e3\}\tabularnewline
\hline
\hline
 \multirow{5}{*}{SVM} & kernel & \{poly, \textbf{rbf}\}\tabularnewline
\cline{2-3}  
 & gamma & \{\textbf{scale}, auto\}\tabularnewline
\cline{2-3} 
 & C & \{1e-2, 1e-1, 1e0, 1e1, \textbf{1e2}\}\tabularnewline
 \cline{2-3} 
 & epsilon & \{1e-3, \textbf{1e-2}, 1e-1, 1e0, 1e1, 1e2, 1e3\}\tabularnewline
 \cline{2-3} 
 & coef0 & \{\textbf{1e0}, 1e-1, 2e-1\}\tabularnewline
\hline
\hline
\multirow{5}{*}{kNN} & algorithm & \{\textbf{ball\_tree}, kd\_tree, brute\}\tabularnewline
\cline{2-3} 
 & n\_neighbors & \{1,2,3,4,5,6,7,8,\textbf{9},10\}\tabularnewline
\cline{2-3}
 & leaf\_size & \{\textbf{1}, 10, 20, 30, 100\}\tabularnewline
 \cline{2-3} 
 & weights & \{uniform, \textbf{distance}\}\tabularnewline
 \cline{2-3}  
 & p & \{\textbf{1}, 2\}\tabularnewline
 \hline
 \hline
\multirow{3}{*}{GP} & n\_restarts\_optimizer & \{(\textbf{0},1,10,100\}\tabularnewline
\cline{2-3}  
 & alpha & \{1e-3, 1e-2, \textbf{1e-1}, 1e0, 1e1, 1e2, 1e3\}\tabularnewline
\cline{2-3} 
 & kernel & \{\textbf{RBF}, ExpSineSquared, RationalQuadratic, Matern\}
 \tabularnewline
\hline
\hline
 \multirow{3}{*}{DT} & criterion & \{\textbf{mse}, friedman\_mse, mae\}\tabularnewline
\cline{2-3}  
 & splitter & \{\textbf{best}, random\}\tabularnewline
\cline{2-3}  
 & max\_features & \{\textbf{auto}, sqrt, log2\}\tabularnewline
 \hline
 \hline
\multirow{8}{*}{RF} & n\_estimators & \{10, \textbf{100}, {1000}\}\tabularnewline
\cline{2-3} 
 & min\_weight\_fraction\_leaf & \{0.0, \textbf{0.1}, 0.2, 0.3, 0.4, 0.5\}\tabularnewline
\cline{2-3} 
 & max\_features & \{sqrt, log2, {\textbf{auto}}\}\tabularnewline
 \cline{2-3}  
 & criterion & \{\textbf{mse}, mae\}\tabularnewline
 \cline{2-3}  
 & min\_samples\_leaf & \{1, 2 ,3, 4, 5, \textbf{10}, 100\}\tabularnewline
 \cline{2-3}  
 & bootstrap & \{\textbf{True}, False\}\tabularnewline
 \cline{2-3} 
 & warm\_start & \{True, \textbf{False}\}\tabularnewline
 \cline{2-3}  
 & max\_impurity\_decrease & \{0.1, \textbf{0.2}, 0.3, 0.4, 0.5\}\tabularnewline
 
\hline 
\hline
\multirow{6}{*}{ET} & n\_estimators & \{10, 100, {\textbf{1000}}\}\tabularnewline
\cline{2-3} 
 & min\_weight\_fraction\_leaf & \{{\textbf{0.0}}, 0.25, 0.5\}\tabularnewline
\cline{2-3} 
 & max\_depth & \{1, 10, 100, {\textbf{None}}\}\tabularnewline
\cline{2-3} 
 & max\_leaf\_nodes & \{2, 10, \textbf{100}\}\tabularnewline
\cline{2-3} 
 & min\_samples\_split & \{2, \textbf{10}, 100\}\tabularnewline
\cline{2-3}  
 & min\_samples\_leaf & \{1, 10, \textbf{100}\}\tabularnewline
\hline 
\hline
\multirow{9}{*}{GB} & n\_estimators & \{10, \textbf{100}, {1000}\}\tabularnewline
\cline{2-3} 
 & min\_weight\_fraction\_leaf & \{\textbf{0.0}, 0.1, 0.2, 0.3, 0.4, 0.5\}\tabularnewline
 \cline{2-3}  
 & max\_features & \{sqrt, log2, \textbf{auto}, None\}\tabularnewline
 \cline{2-3}  
 & warm\_start & \{True, \textbf{False}\}\tabularnewline
 \cline{2-3}  
 & max\_depth & \{1, 10, 100, \textbf{None}\}\tabularnewline
 \cline{2-3} 
 & criterion & \{friedman\_mse, mse, \textbf{mae}\}\tabularnewline
\cline{2-3}
 & min\_samples\_split & \{\textbf{2}, {5}, 10\}\tabularnewline
\cline{2-3} 
 & min\_samples\_leaf & \{1, \textbf{10}, 100\}\tabularnewline
 \cline{2-3} 
 & loss & \{\textbf{ls}, lad, huber, quantile\}\tabularnewline
\hline 
\hline
\multirow{3}{*}{HGB} & loss & \{\textbf{least\_squares}, least\_absolute\_deviation, poisson\}\tabularnewline
\cline{2-3} 
 & min\_sample\_leaf & \{1, 5, 10, 15, 20, \textbf{25}, 50, 100\}\tabularnewline
\cline{2-3} 
 & warm\_start & \{\textbf{True}, False\}\tabularnewline
\hline 
\hline
\multirow{8}{*}{MLP} & activation & \{\textbf{tanh}, relu\}\tabularnewline
\cline{2-3} 
 & hidden\_layer\_sizes & \{10, 50, 100, \textbf{150}, 200\}\tabularnewline
\cline{2-3} 
 & solver & \{\textbf{lbfgs}, adam, sgd\}\tabularnewline
\cline{2-3}
 & leaning\_rate & \{constant, invscaling, \textbf{adaptive}\}\tabularnewline
 \cline{2-3} 
 & nesterovs\_momentum & \{\textbf{True}, False\}\tabularnewline
  \cline{2-3} 
 & warm\_start & \{True, \textbf{False}\}\tabularnewline
  \cline{2-3} 
 & early\_stopping & \{\textbf{True}, False\}\tabularnewline
  \cline{2-3} 
 & alpha & \{0.00001, \textbf{0.0001}, 0.001, 0.01, 0.1, 0.0\}\tabularnewline
\hline
\hline
\end{tabular}
}
\par\end{centering}
\end{table}
\section{Machine learning coupled with ODE solver}
\label{sec:coupling}
In the previous section, several ML methods for the regression of the relaxation terms, Eq.~\ref{eq:-36}, defined in the framework of the STS formulation, have been compared. It was found that by an appropriate selection of hyperparameters, for example, through a cross-validation technique, satisfactorily accurate predictions can be achieved in shorter times respect to traditional methods. A further continuation of the previous task, may then consist in exploiting the potential of the ML to alleviate the computational cost of kinetic processes. 

In the present section, an interface between the best-performing ML algorithm and an ODE solver is explored. Specifically, the same code has been implemented in Matlab and Fortran. In~\cite{campoli2020models} a further comparison between the two implementations was provided. The baseline solution and dataset were generated by running the Matlab version, as described in the previous section, until the equilibrium was reached. The best-performing ML algorithm, previously trained, tested and validated was deployed as a \texttt{pickle} module to be fed with input data from the solver.

\subsection{Matlab-Python interface}
On the application side, direct Python call functionality from Matlab is used. It is possible, in fact, to access Python libraries, functions or classes from Matlab by adding the \texttt{py} prefix to the Python name, as shown in Listing~\ref{lst:runregressor} which calls the ML regressor model, reported in Listing~\ref{lst:regressor} which simply loads scalers, reshapes and transforms the input variable array and performs the prediction. 

A one-dimensional reactive shock flow relaxation in the framework of STS formulation is considered. Further details about this test case and results can be found in~\cite{campoli2020models}. From the computational point of view, the problem reduces to the integration of an ODE system up to the equilibrium state within selected relative and absolute error tolerances, as shown in Listing~\ref{lst:ode}. Due to the stiff nature of the test case, \texttt{ode15s} is used which, in turn, calls the \texttt{rpart} function which is responsible for the computation of the right-hand side of the system of equations, reported in Listing~\ref{lst:rpart}. It is worth observing that Listing~\ref{lst:ode} presents the solver code for the binary mixture. Nevertheless, it can be easily modified to consider air mixtures.

A first aspect to consider is where actually apply the ML. To the best of the authors' knowledge, in fact, this aspect has not been fully detailed in literature. Considering the Listing~\ref{lst:ode} and ~\ref{lst:rpart}, at least four places appear to be possible candidates and correspondingly the regression of different targets can be performed:

\begin{enumerate}
    \item regression of chemical reaction rate coefficients, $k_{ci}$, Eq.~\ref{eq:-30}-\ref{eq:-32} (lines 56-81 of Listing~\ref{lst:rpart});
    \item regression of chemical reaction relaxation terms, $R_{ci}$, Eq.~\ref{eq:-36} (lines 87-120 of Listing~\ref{lst:rpart}, before matrix inversion at line 129);
    \item regression of the right-hand side inside ODE function call, $dy$ (after matrix inversion at line 129 of Listing~\ref{lst:rpart});
    \item regression of the ODE solver function call output, [X,Y] at line 2 of Listing~\ref{lst:ode}).
\end{enumerate}

Option (1) would be certainly possible due to the simple temperature dependence of the rate coefficients which would make their regression quite straightforward. Nevertheless, it would provide a minimal speed-up as we should still perform the expensive main loop to compute the relaxation terms, $R_{ci}$, (lines 88-121 of Listing~\ref{lst:rpart}) and a non-negligible communication time would be required for the Python function calls within the loop itself. This option, then, was not further investigated but it may be a reasonable choice depending on the problem's features. 

\begin{figure}[H]
\centering
\includegraphics[scale=0.4]{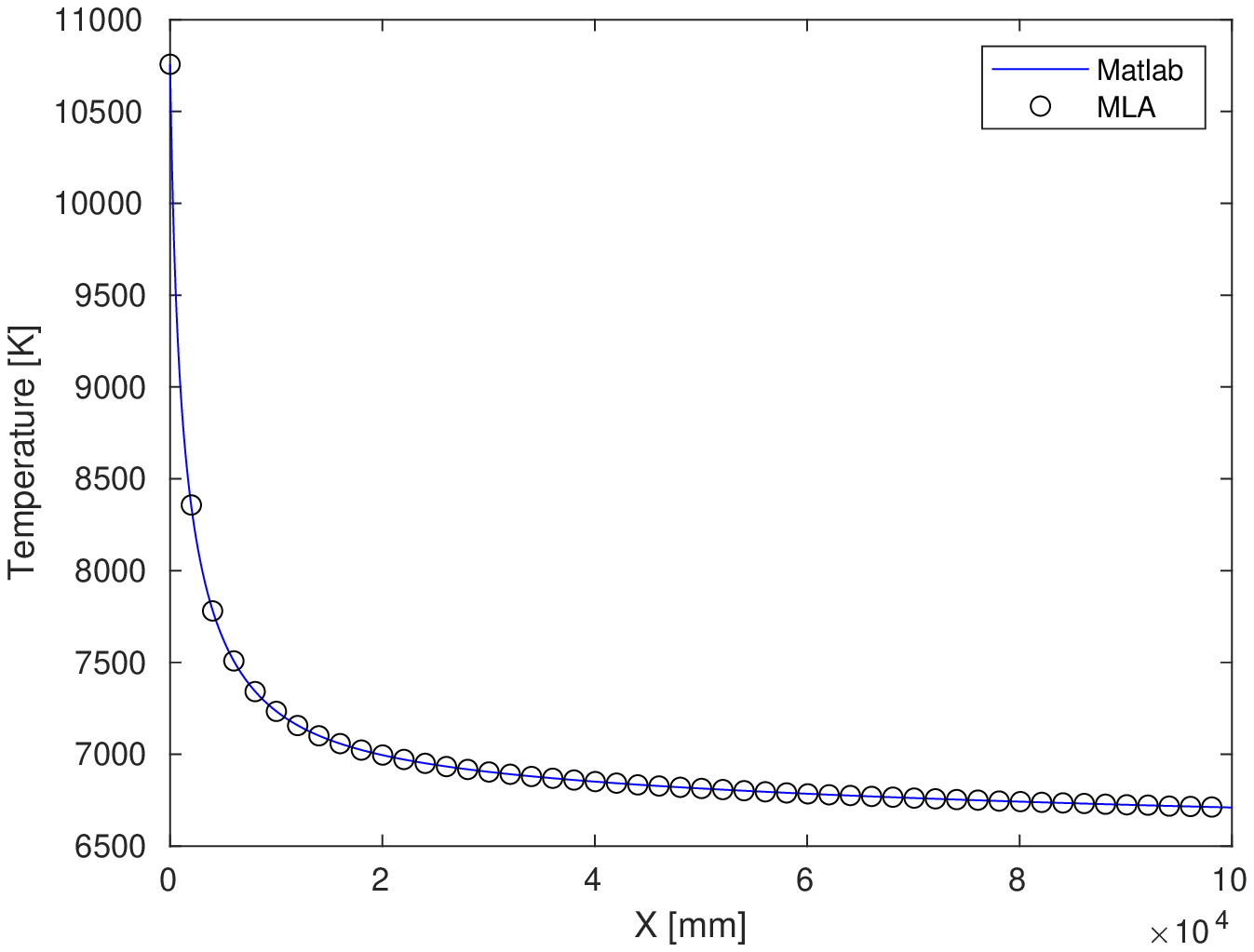}
\includegraphics[scale=0.4]{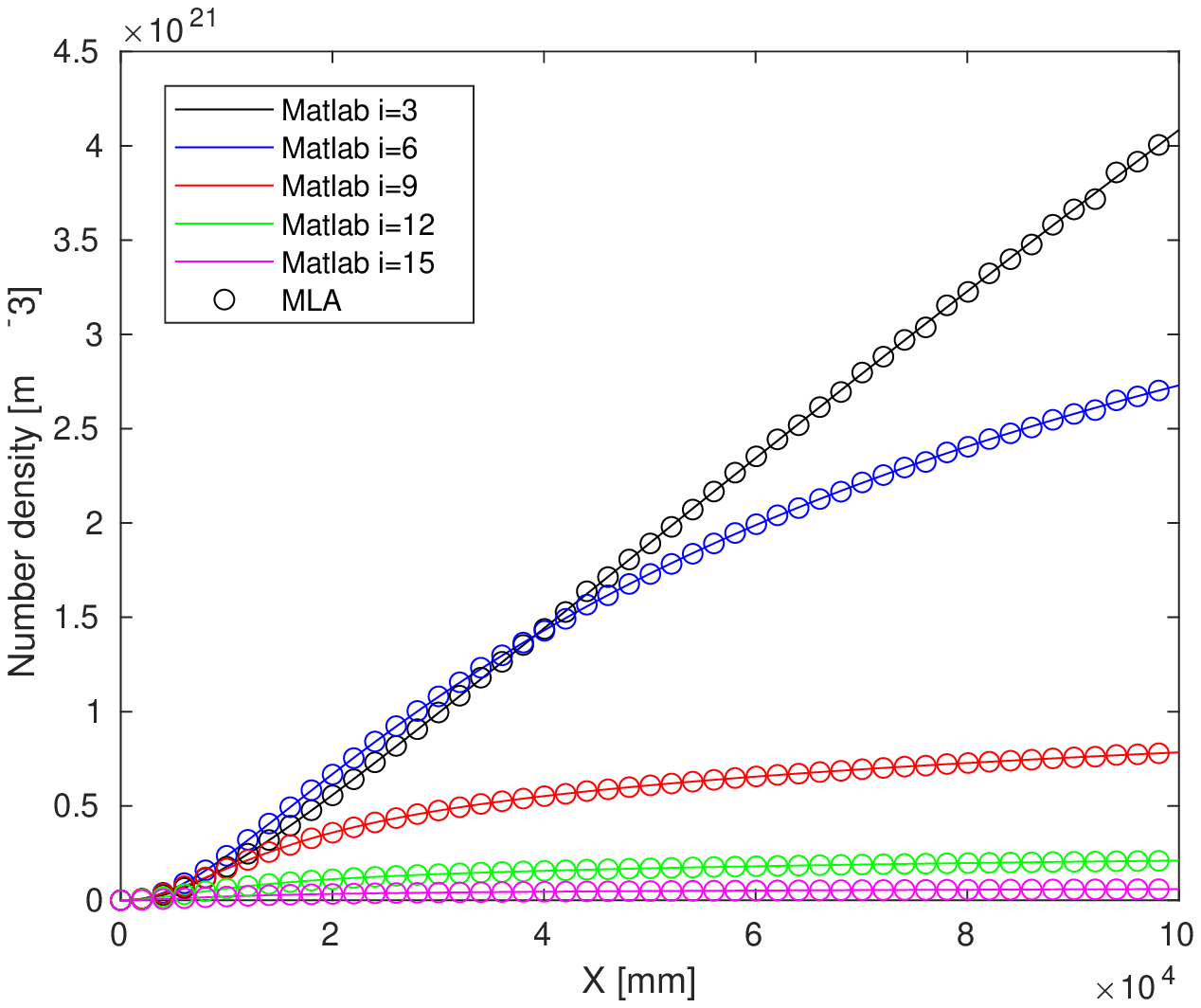}
\caption{Comparison of Matlab and ML solution for the one-dimensional reacting shock flow in STS approach for binary N$_2$/N mixture. 
}
\label{fig:Matlab_mla_out}
\end{figure}

\begin{figure}[H]
\centering
\includegraphics[scale=0.4]{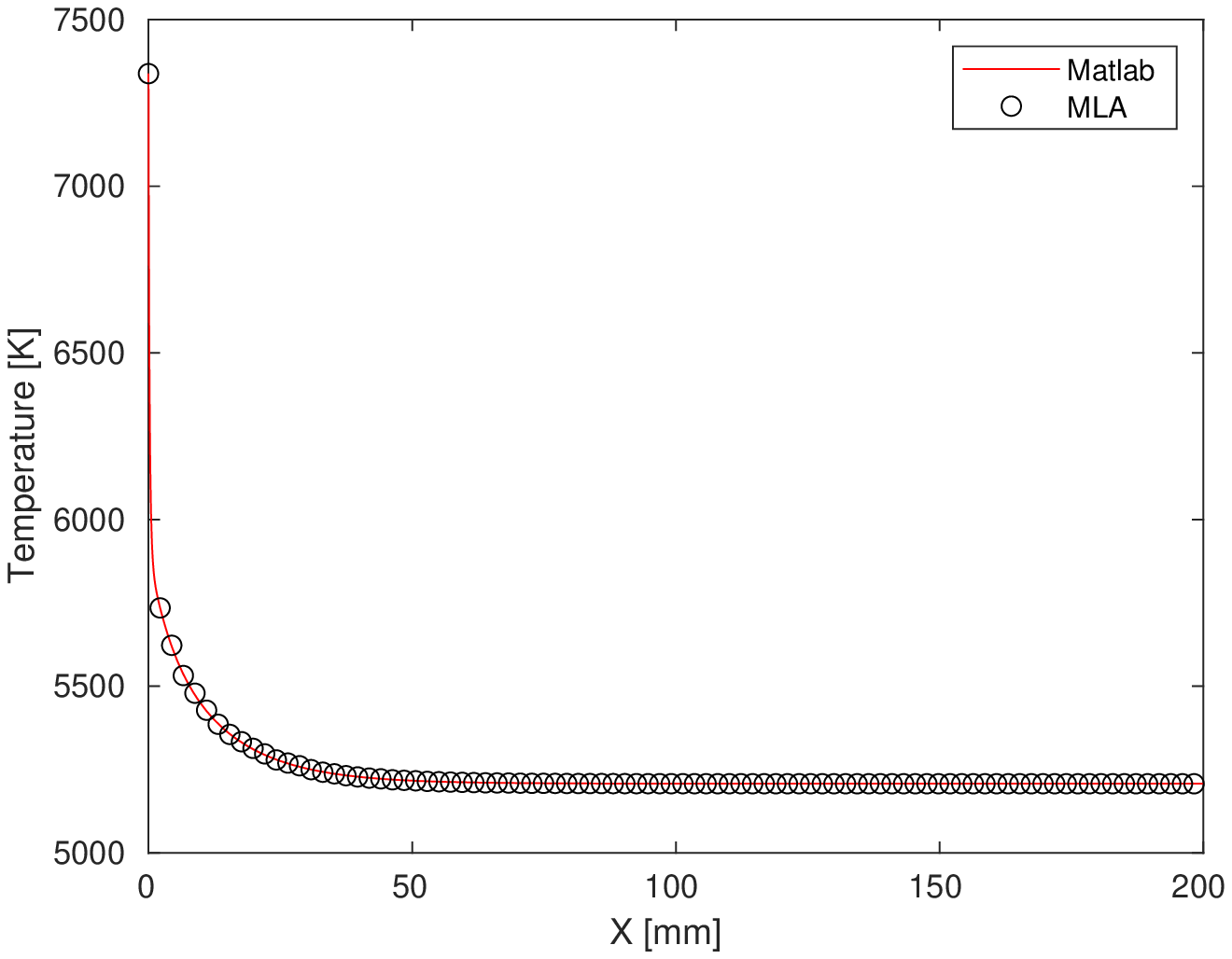}
\includegraphics[scale=0.4]{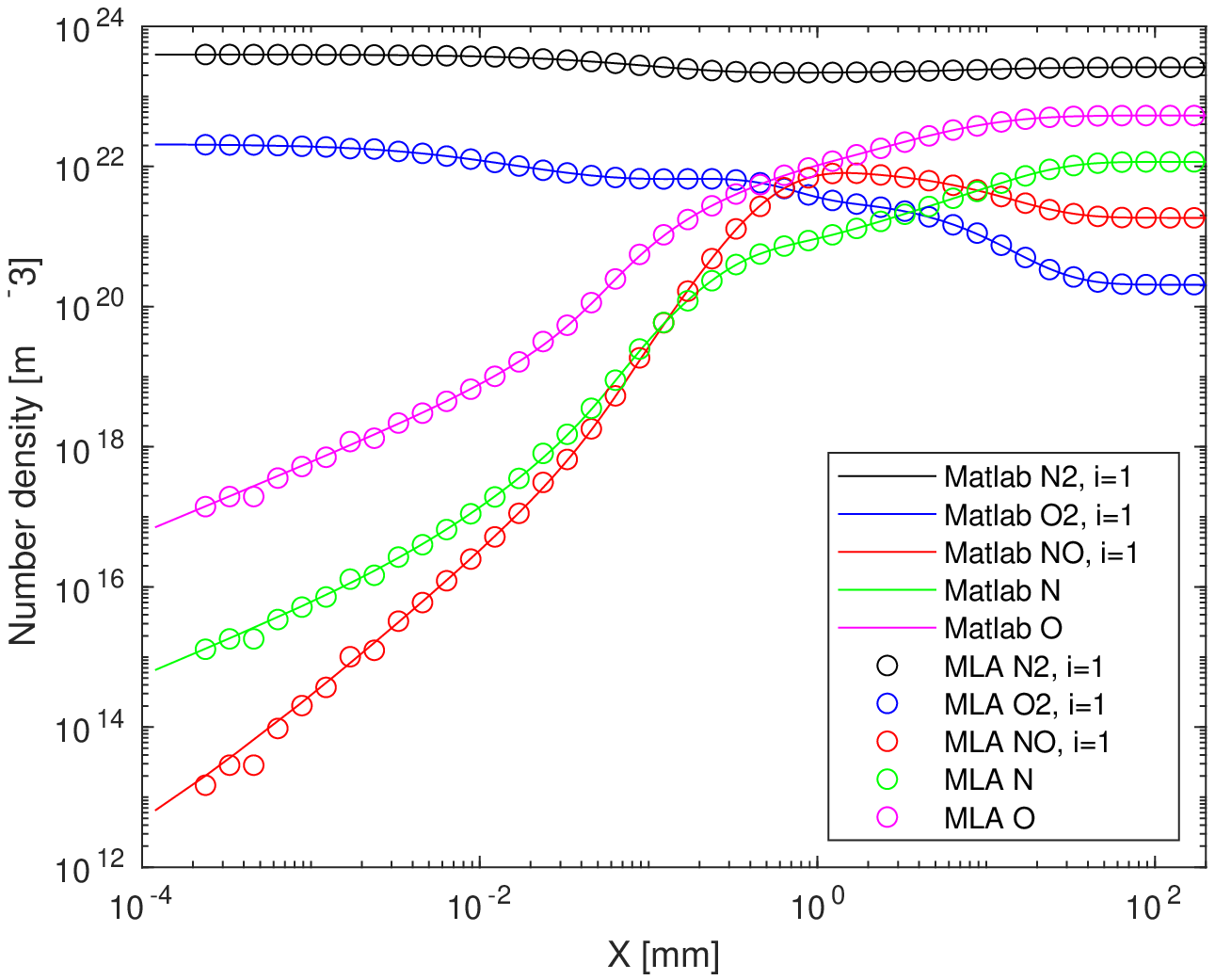}
\caption{Comparison of Matlab and ML solution for the one-dimensional reacting shock flow in STS approach for Air 5 mixture.
}
\label{fig:Matlab_mla_air5_in}
\end{figure}

Option (4), to learn and predict the output of the ODE solver, would allow us to circumvent the call to the integrator, \textit{tout court}, providing the greatest speed-up respect to the baseline solution. In this case, the distance from the shock wave (or equivalently, the relaxation time) was employed as input feature while the species number density (molecule vibrational levels), velocity and temperature were predicted. Figure~\ref{fig:Matlab_mla_out} and~\ref{fig:Matlab_mla_air5_in} report the profile of temperature and number density for selected vibrational levels obtained with this approach by Matlab and ML for binary and air mixture. Satisfactory agreement was obtained in both cases. 

Another relevant aspect to note is the computational cost. Table~\ref{tab:timing} shows the time-to-solution obtained with Matlab and ML for binary and air mixture. As expected, there is no appreciable gain in using ML for simple binary mixtures for whom traditional methods perform well. Nevertheless, it is also worth mentioning here that for N$_2$/N, a computationally simple SSH model is used whereas for air5, a much more expensive FHO model was adopted. Thus, by using FHO for N$_2$/N, an appreciable speed-up will be noticeable. 
Moreover, it can be observed a significant speed-up ($\sim$ 300x) of the solution for more complex mixtures. Moreover, the computational cost (CPU time) of the ML is almost independent on the number of kinetic processes taking place  which means that the more complex mixtures, the more significant gain will be obtainable with ML respect to the Matlab baseline. Furthermore, the CPU time does not depend on the timestep nor on the local stiffness of the ODE system and, most importantly, the storage requirements are expected to grow only moderately as the number of input scalars is increased~\cite{blasco1999single}. These characteristics appear to be quite appealing in the framework of STS approaches, which tend to saturate the computational resources with bottlenecks associated to chemical (and transport) processes.

\begin{lstlisting}[language=Matlab, caption=Matlab function call to the stiff ODE solver, label={lst:ode}]
options = odeset('RelTol', 1e-12, 'AbsTol', 1e-12);
[X,Y] = ode15s(@rpart, xspan, Y0_bar,options);
\end{lstlisting}
\begin{lstlisting}[language=Matlab, caption=Matlab function for source term calculation, label={lst:rpart}]
function dy = rpart(t,y)

format long e
global c h k m l e_i e_0 Be D n0 v0 T0 Delta

Lmax = l-1;

ni_b = y(1:l);
na_b = y(l+1);
nm_b = sum(ni_b);
v_b = y(l+2);
T_b = y(l+3);
temp = T_b*T0;

ef_b = 0.5*D/T0;
ei_b = e_i/(k*T0);
e0_b = e_0/(k*T0);

sigma = 2;
Theta_r = Be*h*c/k;
Z_rot = temp./(sigma.*Theta_r);

M = sum(m);
mb = m/M; 

% A*X=B
A = zeros(l+3,l+3);

for i=1:l
    A(i,i) = v_b;
    A(i,l+2) = ni_b(i);
end

A(l+1,l+1) = v_b;
A(l+1,l+2) = na_b;

for i=1:l+1
    A(l+2,i) = T_b;
end

A(l+2,l+2) = M*v0^2/k/T0*(mb(1)*nm_b+mb(2)*na_b)*v_b;
A(l+2,l+3) = nm_b+na_b;

for i=1:l
    A(l+3,i) = 2.5*T_b+ei_b(i)+e0_b;
end

A(l+3,l+1) = 1.5*T_b+ef_b;
A(l+3,l+2) = 1/v_b*(3.5*nm_b*T_b+2.5*na_b*T_b + ...
             sum((ei_b+e0_b).*ni_b)+ef_b*na_b);
A(l+3,l+3) = 2.5*nm_b+1.5*na_b;

AA = sparse(A);

B = zeros(l+3,1);

Kdr = (m(1)*h^2/(m(2)*m(2)*2*pi*k*temp))^(3/2)*Z_rot*...
      exp(-e_i'/(k*temp))*exp(D/temp);

Kvt = exp((e_i(1:end-1)-e_i(2:end))/(k*temp))';

kd = kdis(temp) * Delta*n0/v0;

kr = zeros(2,l);
for iM = 1:2
    kr(iM,:) = kd(iM,:) .* Kdr * n0;
end

% VT: i+1 -> i
kvt_down = kvt_ssh(temp) * Delta*n0/v0;
kvt_up = zeros(2,Lmax);
for ip = 1:2
    kvt_up(ip,:) = kvt_down(ip,:) .* Kvt;
end

% VV
kvv_down = kvv_ssh(temp) * Delta*n0/v0;
kvv_up = zeros(Lmax,Lmax);
deps = e_i(1:end-1)-e_i(2:end);
for ip = 1:Lmax
    kvv_up(ip,:) = kvv_down(ip,:) .* exp((deps(ip)-deps') / (k*temp));
end

RD  = zeros(l,1);
RVT = zeros(l,1);
RVV = zeros(l,1);

for i1 = 1:l

    RD(i1) = nm_b*(na_b*na_b*kr(1,i1)-ni_b(i1)*kd(1,i1)) + ...
             na_b*(na_b*na_b*kr(2,i1)-ni_b(i1)*kd(2,i1));

     if i1 == 1 %  0<->1
 
         RVT(i1) = nm_b*(ni_b(i1+1)*kvt_down(1,i1) - ni_b(i1)*kvt_up(1,i1))+...
                   na_b*(ni_b(i1+1)*kvt_down(2,i1) - ni_b(i1)*kvt_up(2,i1));
 
         RVV(i1) = ni_b(i1+1)*sum(ni_b(1:end-1) .* kvv_down(i1,:)') - ...
                   ni_b(i1)  *sum(ni_b(2:end)   .* kvv_up(i1,:)');
 
     elseif i1 == l % Lmax <-> Lmax-1
 
         RVT(i1) = nm_b*(ni_b(i1-1)*kvt_up(1,i1-1) - ni_b(i1)*kvt_down(1,i1-1))+...
                   na_b*(ni_b(i1-1)*kvt_up(2,i1-1) - ni_b(i1)*kvt_down(2,i1-1));
 
         RVV(i1) = ni_b(i1-1)*sum(ni_b(2:end)   .* kvv_up(i1-1,:)') - ...
                   ni_b(i1)  *sum(ni_b(1:end-1) .* kvv_down(i1-1,:)');
 
     else
 
         RVT(i1) = nm_b*(ni_b(i1+1)*kvt_down(1,i1)+ni_b(i1-1)*kvt_up(1,i1-1)-...
                   ni_b(i1)*(kvt_up(1,i1)+kvt_down(1,i1-1)))+...
                   na_b*(ni_b(i1+1)*kvt_down(2,i1)+ni_b(i1-1)*kvt_up(2,i1-1)-...
                   ni_b(i1)*(kvt_up(2,i1)+kvt_down(2,i1-1)));
 
         RVV(i1) = ni_b(i1+1)*sum(ni_b(1:end-1) .* kvv_down(i1,:)') + ...
                   ni_b(i1-1)*sum(ni_b(2:end)   .* kvv_up(i1-1,:)') - ...
                   ni_b(i1) *(sum(ni_b(2:end)   .* kvv_up(i1,:)') + ...
                              sum(ni_b(1:end-1) .* kvv_down(i1-1,:)'));
     end
end

B(1:l) = RD + RVT + RVV;
B(l+1) = - 2*sum(RD);

dy = AA^(-1)*B;
\end{lstlisting}

\begin{lstlisting}[language=Matlab, caption=Matlab function call to Python regressor, label={lst:runregressor}]
for i = 1:nsteps 
    input = my_xspan(i);
    RHS = py.run_regression.regressor(input);
    RHS = double(RHS);
end
\end{lstlisting}

\begin{lstlisting}[language=Python, caption=ML regressor, label={lst:regressor}]
import numpy as np
import joblib

def regressor(input):

    # Load scalers
    sc_x = load(open('scaler_x.pkl', 'rb'))
    sc_y = load(open('scaler_y.pkl', 'rb'))

    # Load model
    regr = load('model.pkl')

    # Build array of inputs for prediction
    Xinput = np.asarray(input).reshape(-1,1)

    # Scaler input arguments
    Xinput = sc_x.transform(Xinput)

    # Prediction
    y_regr = regr.predict(Xinput)

    # Inverse transformation
    y_regr_dim  = sc_y.inverse_transform(y_regr)

    return y_regr_dim
\end{lstlisting}

\begin{figure}[H]
\centering
\includegraphics[scale=0.3]{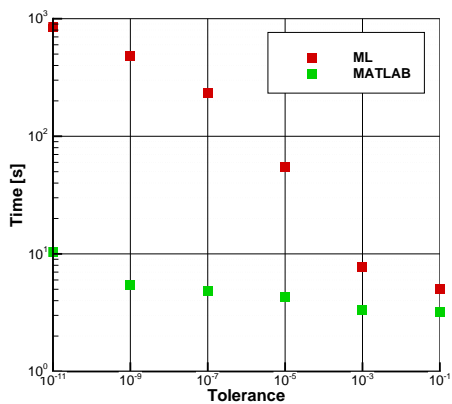}
\caption{Comparison of Matlab and ML time-to-solution for the one-dimensional reacting shock flow in STS approach for binary N$_2$/N mixture. The ML call is performed within the ODE system integration before the matrix inversion.}
\label{fig:Matlab_mla_time}
\end{figure}

\begin{table}[H]
\caption{Comparison of time-to-simulation for Matlab and ML solutions for the same number of integration points.}
\label{tab:timing}
\begin{centering}
\begin{tabular}{cccccccc}
\hline 
& & \multicolumn{3}{c}{N$_2$/N} & \multicolumn{2}{c}{Air 5}\tabularnewline
\hline 
\hline 
 & \multicolumn{2}{c}{Matlab} & ML & FANN & Matlab & ML & FANN \tabularnewline
\hline 
Time {[}s{]} & \multicolumn{2}{c}{7.3541} & 6.8475 & 0.09 & 1874.7 & 6.8974 & 0.11 \tabularnewline
\hline 
\hline 
\end{tabular}
\par\end{centering}
\end{table}
Options (2) and (3), to learn the relaxation terms before or after the A matrix inversion, would both permit to avoid the main loop (lines 88-121 of Listing~\ref{lst:rpart}) but evidently, option (3) will be faster by-passing the A matrix computation and inversion at each step.
Nevertheless, when trying to apply this option (2 or 3, equivalently), we obtain results shown in Figure~\ref{fig:Matlab_mla_time}. The ML solution time tends to diverge as soon as the tolerance is decreased.
This behaviour is connected to the nature of the solution methods for initial value problems (IVPs) which makes the usage of ML difficult when applied to the primary state variables but suitable for secondary property prediction. The nature of boundary value problems (BVPs) makes them easier to hybridize as the field can be predicted by ML and corrected to a defined tolerance more easily. 

{Stiff chemistry solvers, in fact, fall into the class of IVPs while most other problems in CFD are BVPs. This distinction turned out to be quite important in the effort to couple Matlab ODE solver with ML. The problem with using ML to predict the integration of the relaxation terms is that the accuracy of the predicted values is not sufficient when such values are repeatedly fed into the ML model. Even with relative prediction errors reaching as low as $10^-5$, the solver solution slowly diverges from a physically meaningful value. While it is true that evaluating the ML based predictions are exceptionally quick, the nature of the IVP does not allow for efficient correction of the ML predictions as observed also in~\cite{buchheit2019stabilized,kelp2018orders}.}

The complementary nature between machine learning models and differential equations has been recently noted in~\cite{chen2018neural,rackauckas2019diffeqflux} where a possible solution was proposed from the perspective of neural ODE, not further investigated in the present paper.

\subsection{Fortran-Python interface}
Python has de-facto become the lingua-franca of the machine learning community. Numerous packages like Scikit-Learn~\cite{pedregosa2011scikit}, TensorFlow~\cite{abadi2016tensorflow}, PyTorch~\cite{paszke2019pytorch}, Caffe~\cite{jia2014caffe}, MXNet~\cite{chen2015mxnet}, Theano~\cite{al2016theano}, even if written in C/C++/CUDA, they always rely on a Python interface. On the other hand, C/C++ and Fortran still represent the mainstream languages for HPC applications. While the realization of a Matlab-Python interface is straightforward, integrating trained ML components back into Fortran-based  codes is not trivial~\cite{wang2005general} and well documented. In this regards, as we plan to undertake such a task, we investigated existing options.  

There are, in fact, several ways to interface ML frameworks into a Fortran code:
\begin{itemize}
    \item Re-coding specific model architectures into Fortran
    \item Calling Python from within Fortran (e.g. using wrapper libraries such as Python's C API~\cite{van2002python}, Cython~\cite{behnel2011cython}, CFFI\footnote{https://cffi.readthedocs.io}, SWIG~\cite{beazley1996swig}\footnote{http://www.swig.org/projects.html}, Babel~\cite{prantl2011interfacing}, SIP\footnote{https://riverbankcomputing.com/software/sip/intro} or Boost.python library\footnote{https://wiki.python.org/moin/boost.python})
    \item Use a pure Fortran NN library, or bridging library (there are several existing solutions e.g. FANN\footnote{https://github.com/libfann}, neural-fortran\footnote{https://github.com/modern-fortran/neural-fortran}, FKB\footnote{https://github.com/scientific-computing/FKB}, frugally-deep\footnote{https://github.com/Dobiasd/frugally-deep}, Ro-boDNN\cite{szemenyei2018real}, TensorflowLite~\cite{abadi2016tensorflow} C/C++ API and tiny-dnn\footnote{https://github.com/tiny-dnn/tiny-dnn})
    \item Intrinsic Fortran procedures, such as \texttt{get\_command\_argument}, \texttt{get\_command} to invoke Python scripts and exchange data through I/O files.
\end{itemize}

The first option would probably provide efficient solutions and good compatibility with existing Fortran codes. Nevertheless, it would be time-consuming and inflexible as changing the ML model architecture means recoding in Fortran. The last option is the easiest and the less efficient and certainly not adequate for HPC frameworks. 

Pure Fortran or bridging libraries, would can provide fast solutions and good compatibility with Fortran codes. At the present moment, the existing solutions only have limited architectures and algorithms available since they are either Fortran re-implementations of methods or APIs that mirror common ML frameworks. Translations from native ML model format are often required. The case of bridging libraries in C/C++ would introduce again the necessity to have an additional interface layer.

Using wrappers is by far one of the most frequent solutions~\cite{johnson2019automated,prokopenko2019documenting,evans2017existing,young2017fortrilinos}. In this case, no re-coding is required as the ML model remains in Python but various compatibility issues and technicalities may arise depending on the wrapping approach. 

Finally, the direct interface to C/C++ ML frameworks (i.e. TensorFlow) may be an interesting option. The primary benefit to this approach is flexibility: the implementation can be changed, extended, or optimized without affecting the code integration, the library can be integrated into new codes without requiring a complicated extraction, and the code runs on multiple types of hardware and performs lightweight inference without requiring the full TensorFlow ecosystem.

Considering, for example, the \texttt{frugally-deep} package, it first converts DNN models to json files, and then provides C++ header classes that allow loading of json files as object graphs that can be evaluated on the input data. 

After installation, a typical workflow would be as follows:
\begin{itemize}
    \item Create, train and save deep learning model from Python:\\ \texttt{model.save("keras\_model.h5", include\_optimizer=False)}
    \item Convert the saved model into the required format:\\
    \texttt{python3 convert\_model.py keras\_model.h5 fdeep\_model.json}
    \item Load model in C++ using \texttt{frugally-deep}:\\
    \texttt{const auto model = fdeep::load\_model("fdeep\_model.json")};
    \item Load data from Fortran
    \begin{itemize}
        \item pass it to function in C++
        \item make inference
        \item pass inference result back to Fortran
    \end{itemize}
\end{itemize}
In order to obtain an estimate of the speed-up attainable by using a direct interface to C/C++ ML frameworks, in the present paper, a simple experiment was conducted. Specifically, the C Fast Artificial Neural Network library (FANN) was binded to the Fortran version of the STS 1D Euler shock relaxation solver. With this configuration, the simulation reported in Tab.~\ref{tab:timing} for the air mixture was repeated. It was found that without any appreciable difference in the qualitative agreement of the results, the time-to-simulation was about 0.1 s, as reported in Tab.~\ref{tab:timing}, that is, 70 times faster that the Matlab/Python interface and about four orders of magnitude faster that the original Matlab solution. This kind of speed-up is in agreement with results found in~\cite{kelp2018orders}.
\section{Deep Neural network for 1D STS Euler shock flow relaxation}
\label{sec:EulerSTS}
In this section, we investigate the possibility of using a deep neural network (DNN) to infer the 1D Euler system of equation's solution for high-speed non-equilibrium reacting flows according to a STS description~\cite{nk_book_e}. We consider the relaxation of a flow across a normal shock wave for a binary N$_2$/N mixture and we shall infer the number density of pseudo-species (47 vibrational levels of N$_2$) and atomic N, $n_{ci}$, density $\rho$, velocity $v$, pressure $p$, specific internal energy  $E$ as well as relaxation source terms $R_{ci}$ for all the considered processes by using the DNN and adopting the distance from the shock front $x$ as input feature descriptor. The dataset is the same as the one described in Sect.~\ref{sec:regression}. Hence, the output vector is made up of 100 variables:
\begin{equation}
\textbf{y} = [n_{ci}, \rho, v, p, E, R_{ci}]
\end{equation}
The dataset was divided by using the built-in \texttt{train\_test\_split} scikit-learn function where 75\% of samples were used for training, whereas the remaining 25\% for testing and successively normalized with \texttt{MinMaxScaler}. For this task, we used a multi-layer perceptron (MLP) architecture, a type of ANN that is well suited for nonlinear regression problems. 
We train the neural network by minimizing average mean-squared-error (MSE) using stochastic gradient descent with two optimizers, an external one by Scipy (L-BFGS-B), a quasi-Newton, full-batch gradient-based optimization algorithm and an internal one by TensorFlow (Adam). Such strategy was found to be beneficial for convergence. The learning rate was kept constant as equal to the default value. We use a limited number of experiments to select the network architecture, batch size and epochs based on performance of the evaluation dataset. It found that this task was not particularly sensible to such parameters as soon as a shallow network was not employed.

In early attempts at training the DNN, the network was trained to simultaneously fit all targets. This was the favored approach because (1) the softmax function could be used as the output activation function and (2) previous researchers have succeeded with this approach. However, it was realized that, due to the non-convex and stiff nature of the optimization problem, the learning algorithm would often get trapped in local minima. It was therefore opted to train individual networks for each target output variable independently. This made the DNN both faster to train and much more accurate, as well as easier to modify. This approach is in line with the work of other authors~\cite{mao2020deepm, cai2020deepm, buchheit2019stabilized, sharma2020deep, kelp2018orders}. 

\begin{figure}[H]
\centering
\includegraphics[scale=0.45]{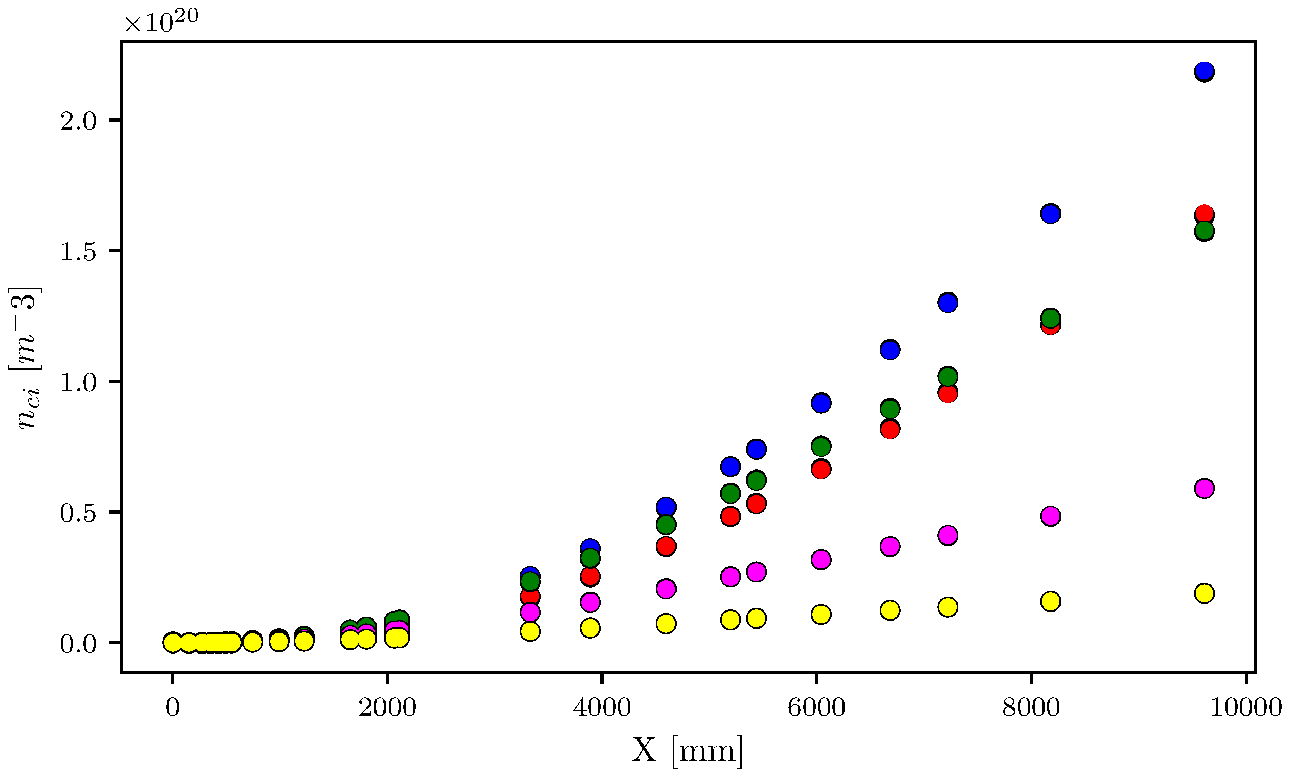}
\includegraphics[scale=0.45]{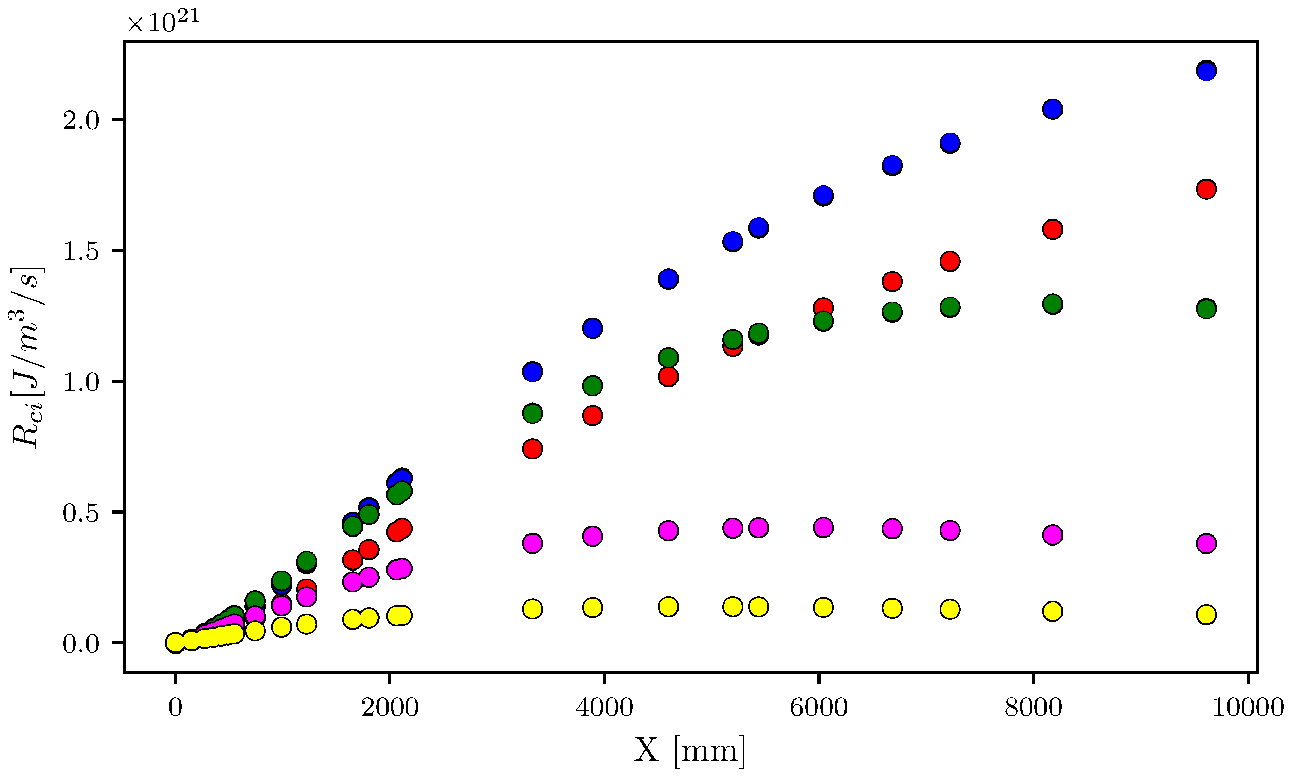}
\\
\includegraphics[scale=0.44]{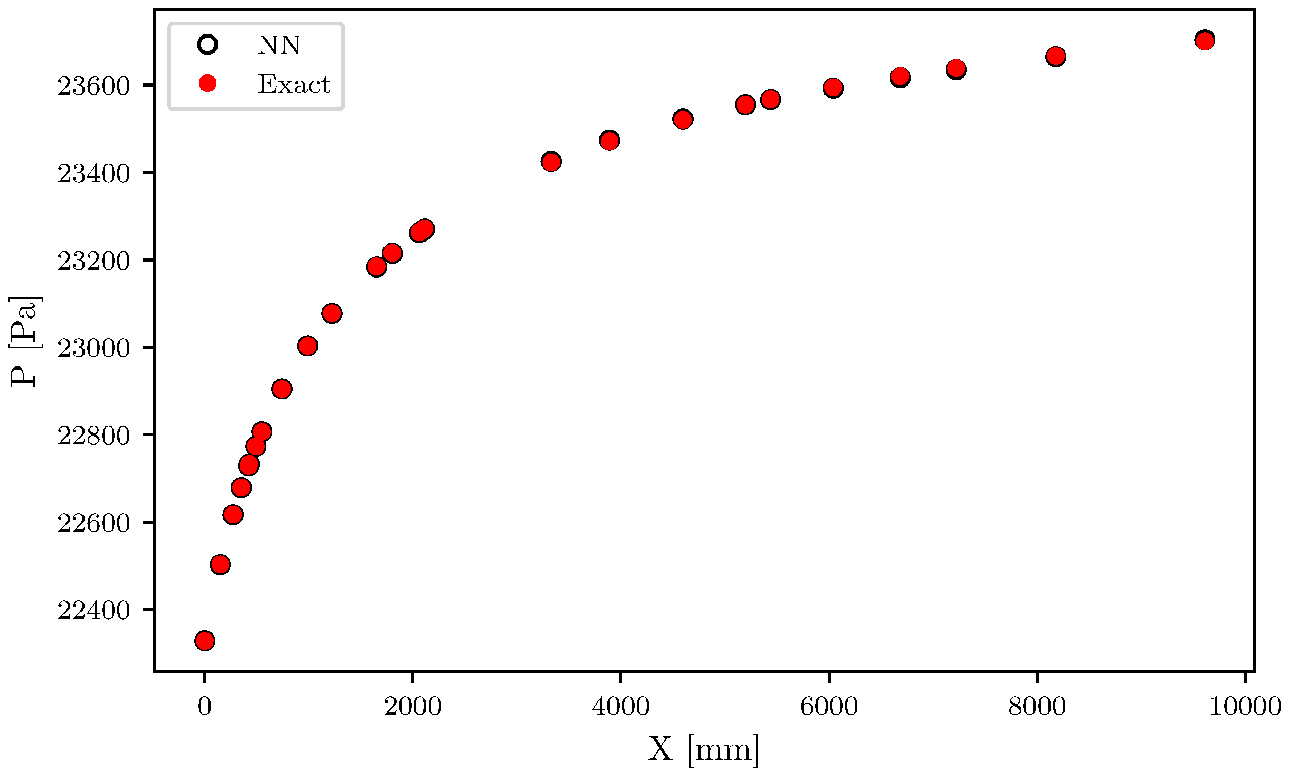}
\includegraphics[scale=0.44]{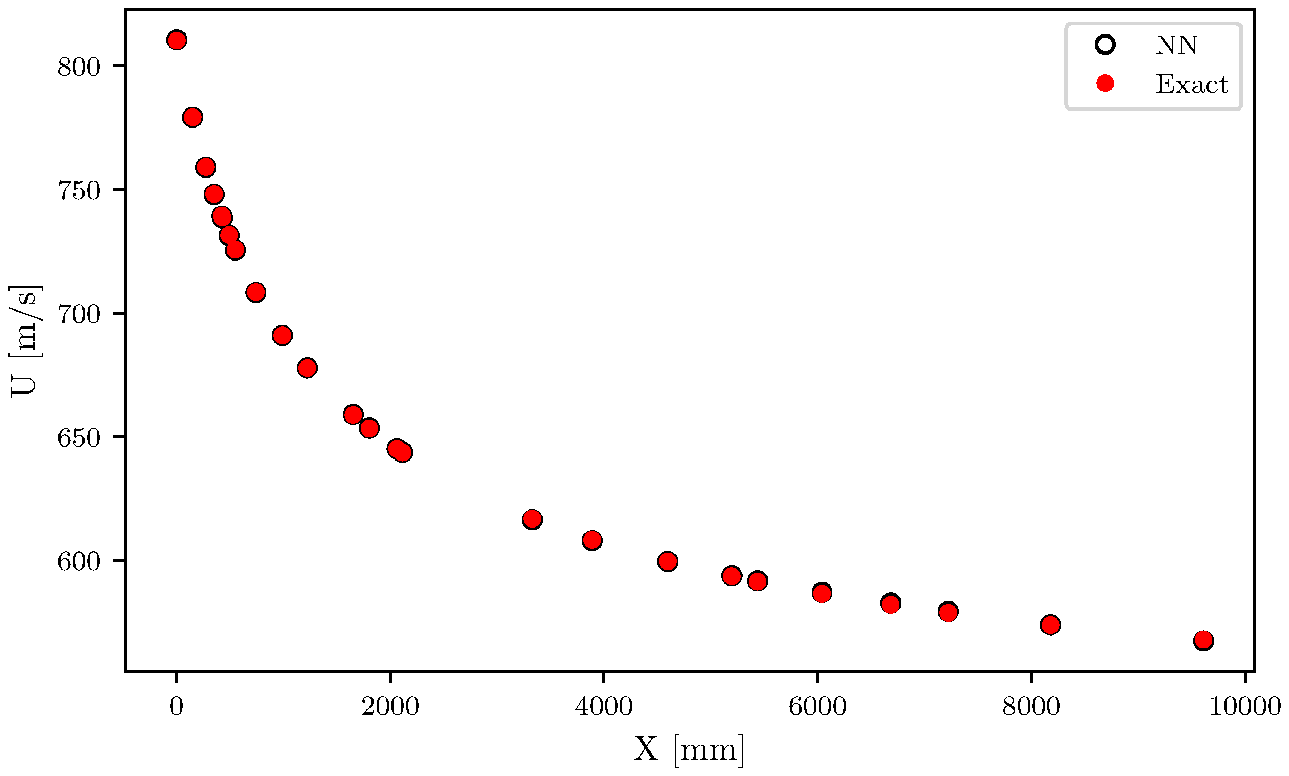}
\caption{Solution of the STS 1D shock flow relaxation inferred by DNN.}
\label{fig:STS}
\end{figure}

{Figure~\ref{fig:STS} reports the profiles of number density, relaxation rates for few selected vibrational levels, pressure and velocity while Tab.~\ref{tab:MRE} summarizes the mean relative errors. Values inferred by the DNN are compared with the "ground truth".}  Satisfactory agreement was obtained for all targets.

Nevertheless, several open questions remain, for example, regarding the generalization skills and interpretability of such approach. We would like to have a robust DNN able to generalize respect to variations of the full set of initial conditions and able to distinguish the contribution of the different physical processes.

\begin{center}
\begin{table}[H]
\begin{centering}
\caption{Mean Relative Error associated to inferred variables.}
\label{tab:MRE}
\begin{tabular}{cc}
\hline 
Variable & Mean Relative Error\tabularnewline
\hline 
\hline 
$n_{ci}$ $[m^{-3}]$ & 4.890 $\cdot$ $10^{-4}$\tabularnewline
\hline 
$R_{ci}$ $[J/m^3/s]$ & 4.039 $\cdot$ $10^{-4}$\tabularnewline
\hline 
$\rho$ $[kg/m^3]$ & 3.793 $\cdot$ $10^{-4}$\tabularnewline
\hline 
u $[m/s]$ & 3.673 $\cdot$ $10^{-4}$\tabularnewline
\hline 
p [Pa] & 1.083 $\cdot$ $10^{-4}$\tabularnewline
\hline 
E  [eV] & 1.248 $\cdot$ $10^{-4}$\tabularnewline
\hline 
\end{tabular}
\par\end{centering}
\end{table}
\par\end{center}
\section{Conclusions}\label{sec:conclusion}
In this work, we presented an assessment of a subset of machine learning methods to state-to-state formulations applied to a one-dimensional post-shock flow relaxation.

Regression of relaxation terms was first performed and several state-of-the-art ML algorithms were compared. It was found that best performances in terms of prediction time are achieved by Decision Tree algorithm, while minimal error levels are obtained with Kernel Ridge. The k-Nearest Neighbour algorithm provides a good trade-off between prediction time and accuracy. It is also worth noting that even if the performance of the proposed ML framework and the results are found to be satisfactory, nevertheless, accuracy of the ML-based predictions could be further improved with training of bigger dataset and refined hyperparameter tuning/optimization.

Secondly, due to the very small prediction time of the best-performing regressor, a coupling between an ODE solver and ML was attempted. In this case, the aim was to investigate possible speed-up, obtainable by relieving the solver from the heavy computation of the stiff kinetic terms. Several strategies have been discussed and few issues reported.

The third task consisted in directly inferring the full solution of the Euler system of equations for the one-dimensional STS reacting shock flow by exploiting an deep neural network (DNN). In this regards, satisfactory agreement was obtained for all variables of interest. Nevertheless, further research is going on in order to improve generalizability and interpretability.

Fostered by these results, we plan to dig deeper in this direction. In this regard, in fact, several aspects are worth investigating. Machine learning-based approach for simulating hypersonic flows using high-fidelity STS kinetics and transport models, appears to be an interesting technique to address the challenges which inherently emerge in the STS formulation related to the stiffness and the huge number of kinetic terms and to the computational cost of transport coefficients. In a subsequent publication, we plan to extend the present analysis also to the regression of STS transport coefficients and to investigate the possibility of coupling between pre-trained best-performing ML algorithm and CFD solver in order to speed-up also the computation of the transport module for one- and two-dimensional problems.

\section*{Acknowledgements}
This work was supported by the Russian Science Foundation, grant 19-11-00041. 
The authors would like to thank Olga Kunova for kindly providing the Matlab code used in Sect. 4.

\bibliography{refs/refs.bib}

\end{document}